\documentclass[aps,prb,twocolumn,superscriptaddress,floatfix]{revtex4-2}

\usepackage{graphicx}
\usepackage{amssymb, amsmath}
\usepackage{xcolor}
\usepackage{multirow,tabularx}
\usepackage{mathtools}

\usepackage{siunitx}

\usepackage[utf8]{inputenc}

\newcommand{\be}{\begin{equation}}
\newcommand{\ee}{\end{equation}}

\newcommand{\ew}[1]{\left\langle #1\right\rangle}

\newcommand{\ket}[1]{\left|#1\right\rangle}

\newcommand{\Xn}{\text{X}_S^0}
\newcommand{\Xp}{\text{X}_S^+}
\newcommand{\Xm}{\text{X}_S^-}
\newcommand{\XmT}{\text{X}_{S,T}^-}
\newcommand{\XXn}{\text{XX}_S^0}

\newcommand{\XXXn}{\text{XXX}_S^0}

\newcommand{\XmXXn}{\Xm + \XXn}
\newcommand{\Dex}{\Delta_{\text{ex}}}
\newcommand{\Tcorr}{T_{\text{corr}}}
\newcommand{\Tblink}{T_{\text{blink}}}
\newcommand{\AB}{632 nm}
\newcommand{\QRes}{\text{QR}}
\newcommand{\TNR}{T_{\text{NR}}}
\newcommand{\PolDeg}{\text{PolDeg}}
\newcommand{\SpinPrep}{f_\text{SpinPrep}}
\newcommand{\AlGaAs}{\text{Al}_{0.15}\text{Ga}_{0.85}\text{As}}
\newcommand{\DetResp}{\mathcal{N}_\text{Det}(\delta_{\text{det}})}

\newcolumntype{Y}{>{\centering\arraybackslash}X}

\begin{document}


\title{Deterministic preparation of spin qubits in droplet etched GaAs quantum dots using quasi-resonant excitation}

\author{Caspar Hopfmann}
\email[]{c.hopfmann@ifw-dresden.de}

\author{Nand Lal Sharma}

\author{Weijie Nie}
\affiliation{Institute for Integrative Nanosciences, Leibniz IFW Dresden, Helmholtzstraße 20, 01069 Dresden, Germany}


\author{Robert Keil}
\affiliation{Institute for Integrative Nanosciences, Leibniz IFW Dresden, Helmholtzstraße 20, 01069 Dresden, Germany}
\affiliation{Present address: Fraunhofer-Institut für Angewandte Festkörperphysik (IAF), Tullastraße 72, 79108 Freiburg, Germany}

\author{Fei Ding}
\affiliation{Institut für Festkörperphysik, Leibniz Universität Hannover, Appelstraße 2, 30167 Hannover, Germany}

\author{Oliver G. Schmidt}
\affiliation{Institute for Integrative Nanosciences, Leibniz IFW Dresden, Helmholtzstraße 20, 01069 Dresden, Germany}
\affiliation{Material Systems for Nanoelectronics, Technische Universität Chemnitz, 09107 Chemnitz, Germany}
\affiliation{Nanophysics, Faculty of Physics and Würzburg-Dresden Cluster of Excellence ct.qmat, TU Dresden, 01062 Dresden, Germany}


\date{\today}

\begin{abstract}
We present a first comprehensive study on deterministic spin preparation employing excited state resonances of droplet etched GaAs quantum dots. This achievement facilitates future investigations of spin qubit based quantum memories using the GaAs quantum dot material platform. By observation of excitation spectra for a range of fundamental excitonic transitions the properties of different quantum dot energy levels, i.e. shells, are revealed. The innovative use of polarization resolved excitation and detection in quasi-resonant excitation spectroscopy facilitates determination of $85$ $\%$ maximum spin preparation fidelity - irrespective of the relative orientations of lab and quantum dot polarization eigenbases. Additionally, the characteristic non-radiative decay time is investigated as a function of ground state, excitation resonance and excitation power level, yielding decay times as low as $29$ ps for s-p-shell exited state transitions. Finally, by time resolved correlation spectroscopy it is demonstrated that the employed excitation scheme has a significant impact on the electronic environment of quantum dot transitions thereby influencing its charge and coherence.
\end{abstract}


\maketitle

\section{Introduction}

Optically accessible quantum memories are fundamental for practical implementations of quantum networks as they facilitate synchronization required for schemes of long distance quantum information exchange \cite{Cirac1997, Kimble2008,  Simon2017, Loock2019}. While tremendous progress towards optical quantum networks using entangled photon sources has been achieved in the past decade \cite{Ritter2012, Wang2016, Schwartz2016, Liao2018, Zopf2019, BassoBasset2019, Bhaskar2020}, quantum systems that address all of the required aspects - namely: high source efficiency, repetition rates and quantum coherence - are still sorely lacking \cite{Loock2019}. Atomic and diamond defect based systems feature long quantum storage times required for memory applications, their repetition rate and internal quantum efficiency, respectively, are however fundamentally limited \cite{Koerber2018, Abobeih2018}. Due to their rapid technological advance, semiconductor quantum dots (QDs) have also received significant attention \cite{Schwartz2016, Chen2018, Zopf2019, BassoBasset2019}. The main drawback of these systems so far has been that the quantum coherence of their spin qubits is limited by interaction with the nuclear magnetic environment of its constituting atoms - i.e. the Overhauser field  \cite{Bloembergen1948, Gammon2001, Cogan2018}. While nuclear polarization and spin echo techniques may extend the coherence time of spin qubits significantly \cite{Greilich2006a, Hoegele2012}, the achievable values in commonly employed InGaAs QDs are still limited by the presence of high spin (+9/2) $\prescript{115}{}{\text{In}}$ and $\prescript{113}{}{\text{In}}$ isotopes. In recent years droplet etched GaAs QDs have been established as potent sources of entangled photon pairs \cite{Chen2018, Zopf2019, BassoBasset2019, Hopfmann2020}, featuring high quality on-demand entangled photon pair generation at GHz clock rates suitable for entanglement swapping operations. Due to the lack of In, these quantum dots are also attractive candidates for extended spin qubit coherence - which has not been investigated so far. The presented study aims at obtaining the foundational knowledge to facilitate further quantum optical applications and investigations based on GaAs QDs. This first exploration is achieved by demonstrating that quasi-resonant (QR) excitation schemes can be used to initialize fundamental QD spin qubits deterministically. To this end, polarization resolved excitation spectroscopy and time resolved pump-probe investigations are combined to obtain a comprehensive picture of the properties of excited states in GaAs quantum dots and their possible applications. This study therefore provides an ideal stepping stone towards future investigations of quantum memories using GaAs QD spin qubits.

\section{Results and Discussion}

\begin{figure}
	\includegraphics[width=1.0\linewidth]{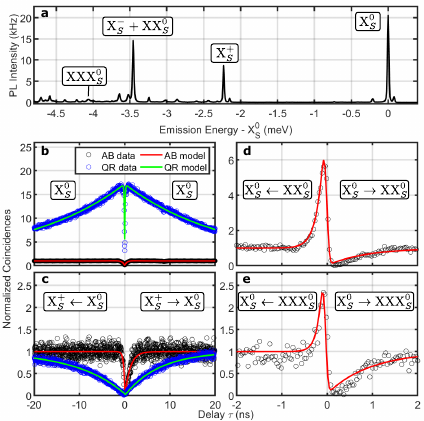}
	
	\caption{(a) Typical spectrum of a single GaAs QD using above-band excitation. Selected fundamental emission lines (c.f. Tab. \ref{tab:lines}) are annotated.  (b) and (c-e) depict $\Xn$ two-photon auto- and cross-correlation traces, respectively, as a function of the time delay $\tau$ between detection events. Traces (fitted model, see text) recorded using above-band (AB) and quasi-resonant (QR, $\Dex = 10.08$ meV, c.f. Fig. \ref{fig:PLE}) excitation are drawn in black (red) and blue (green), respectively. Cross-correlations (c-e) are recorded between $\Xn$ and $\XmXXn$, $\Xp$ as well as $\XXXn$, respectively.}
	\label{fig:basic_PL}
\end{figure}

A typical above-band photoluminescence (PL) spectrum of a single QD is shown if Fig. \ref{fig:basic_PL}(a) in saturation of the neutral QD exciton $\Xn$ transition. Though the general shape of GaAs quantum dot spectra has been established previously \cite{Huber2017}, the exact energy differences between the emission lines depend on the Al concentration of the QD matrix material. Contrary to In(Ga)As-based quantum dots which are grown using the Stranski–Krastanov process, the spectral properties - such as the energetic distribution of fundamental emission lines - of droplet etched GaAs quantum are much more consistent \cite{Keil2017}. In Fig. \ref{fig:basic_PL}(b) transitions of four basic excitonic QD charge configurations, namely $\Xn$, $\Xp$ and $\XmXXn$ are indicated. The transitions were identified using a combination of polarization, power dependent PL and PL excitation (PLE) spectroscopy, details are given in the following sections and in the supplementary material. Note that $\XmXXn$ transition lines are not separable in PL spectra. Using PLE spectroscopy the observed energetic difference between these lines is about $5$ $\mu$eV, which is well within the resolution limit of the employed spectrometer. Spectral overlap between $\Xm$ and $\XXn$ is observed consistently for matrix material Al-concentrations of $15$ $\%$. The nomenclature of fundamental QD transitions used throughout this work is summarized in Tab. \ref{tab:lines}.

\begin{table}
	\begin{tabular}{|c|c|S[table-format=3.2]|}
		\hline
		{Label} & {Transition} & {$\Delta_{i-X_S^0}$ (meV)} \\				
		\hline
		$\Xn$ & $(1e^1)(1h^1) \to 0$ & 0 \\		
		
		$\Xp$ & $(1e^1)(1h^2) \to (1h^1)$ & -2.24\\
		
		$\XXn$ & $(1e^2)(1h^2) \to (1e^1)(1h^1)$ & -3.47 \\
		
		$\Xm$ & $(1e^2)(1h^1) \to (1e^1)$ & -3.48\\
		
		$\XXXn$ & $(1e^2 2e^1)(1h^2 2h^1) \to (1e^1 2e^1)(1h^1 2h^1)$ & -4.09 \\
		\hline
	\end{tabular}
	\caption{Description of various optical transitions of GaAs QDs embedded in $\AlGaAs$ matrix material using the nomenclature established by Benny et. al. \cite{Benny2011a}. Relative emission energies with respect to $\Xn$ ($E_{\Xn} = 1.5880$ eV) are determined by polarization and power dependent high resolution spectroscopy, cf. Fig. \ref{fig:basic_PL}(a) and Suppl. Fig. \ref{fig:Sup_FSS}.} 
	\label{tab:lines}
\end{table}

\subsection{Correlation Spectroscopy}

Figs. \ref{fig:basic_PL}(b-e) depict two-photon auto- and cross-correlation traces $g(\tau)$ using continuous wave (cw) excitation of selected combinations of emission lines. The curves are modeled, depending on the nature of the correlation, according to:

\begin{align}
\label{eq:g2BLK}
g^{(2)}(\tau) & = \left( g_{\text{anti}}(\tau) \times g_{\text{blink}}(\tau) \right) \ast \DetResp \;,  \\
\label{eq:g2RAW}
g_{\text{anti}}^{(2)}(\tau) & =  (g_0-1) \, e^{-\frac{|\tau |}{\Tcorr}}+1 \;,\\
\label{eq:BLK}
g_{\text{blink}}^{(2)}(\tau) & = (\frac{1}{\beta }-1) \, e^{-\frac{\left| \tau \right| }{\Tblink}}+1 \;.
\end{align}

Where $\Tcorr$ and $\Tblink$ are the characteristic correlation and blinking timescales, respectively, and $\beta$ is the blinking on-off ratio. $g_{\text{blink}}(\tau)$ models spectral blinking of the investigated line and is adopted from Jahn et. al. \cite{Jahn2015}. In the limit of an adiabatic (coherent) excitation scheme $T$ is equivalent to the radiative lifetime $\Tcorr \to T_1$ of QD excitonic complexes. In order to take into account the limited temporal response of the detection system the raw auto-correlation trace is convoluted with the bi-detector response function $\DetResp$, modeled by a normal distribution of FWHM $\delta_{\text{det}} = 75$ ps. The latter is determined independently. Eq. (\ref{eq:g2BLK}) can be adopted for cross-correlations by differentiating between regimes $\tau_{i \to j} < 0$ and $\tau_{j \to i} > 0$, effectively yielding two correlation timescales that represent switching between emission events of two different QD excitonic complexes. A summary of the modeled parameter values is shown in Tab. \ref{tab:crossres}. \\ 

\begin{table}
	\begin{tabularx}{\linewidth}{|*{7}{Y|}}
		\hline
		Fig. \ref{fig:basic_PL} &  \multicolumn{2}{c|}{Transition} & \multicolumn{2}{c|}{$g^{(2)}(0)$} & \multicolumn{2}{c|}{$\Tcorr (ns)$} \\
			& $i$ & $j$ & $i \to j$ & $j \to i$ & $i \to j$ & $j \to i$ \\
		\hline
		b(AB)   & \multicolumn{2}{c|}{\multirow{2}{*}{$\Xn$}}	& \multicolumn{2}{c|}{$0.023 \pm 0.006$} 	    & \multicolumn{2}{c|}{$2.15 \pm 0.07$} \\
		b(QR)	& \multicolumn{2}{c|}{} 					  	& \multicolumn{2}{c|}{$0.00 + 0.02$} & \multicolumn{2}{c|}{$0.32 \pm 0.01$} \\
		
		\hline
		c(AB) & \multirow{2}{*}{$\Xn$} & \multirow{2}{*}{$\Xp$} & \multicolumn{2}{c|}{$0.00 + 0.03$} & $0.56 \pm 0.07$ & $2.3 \pm 0.1$ \\
		c(QR) & &												  & \multicolumn{2}{c|}{$0.025 \pm 0.004$} & $21.6 \pm 0.3$ & $15.2 \pm 0.3$ \\
		
		\hline			
		d(AB) & $\Xn$ & $\XXn$ & $0.00 +0.04$ & $9.4 \pm 0.1$ & $0.303 \pm 0.006 $ & $1.70 \pm 0.09$ \\

		\hline	
		e(AB) & $\Xn$ & $\XXXn$ & $0.00 +0.04$ & $3.8 \pm 0.3$ & $0.21 \pm 0.03$ & $1.9 \pm 0.2$ \\
		\hline
		
	\end{tabularx}		
	\caption{Estimated parameter values according to Eq. (\ref{eq:g2BLK}) and (\ref{eq:g2RAW}) of the auto- and cross-correlation traces shown in Fig. \ref{fig:basic_PL}(b-e), respectively. The abbreviations AB and QR stand for above-band and quasi-resonant excitation schemes, respectively. Errors are statistically estimated by $1 \, \sigma$ confidence intervals.} 
	\label{tab:crossres}
\end{table}

Both $\Xn$ auto-correlation traces of Fig. \ref{fig:basic_PL}(b) exhibit strong anti-bunching and are modeled according to Eq.(\ref{eq:g2BLK}). The estimated parameter values related to the blinking are $\beta^{\AB} = 0.86 \pm 0.01$ and $\Tblink^{\AB} = (2.0 \pm 0.1)$ ns, for quasi-resonant excitation values of $\beta^{\QRes} = 0.06 \pm 0.01$  and $\Tblink^{\QRes} = (22.3 \pm 0.3)$ ns are obtained. While the $g^{(2)}(0)$ values for both excitation schemes are comparable, the modeled timescales are very divergent. The difference in $\Tcorr$ can be readily understood when considering that the quasi-resonant excitation is a significantly more coherent excitation scheme than above-band pumping, in fact $\Tcorr^{\QRes} \to T_1$ holds as confirmed below. This is a first indication that quasi-resonant excitation schemes in GaAs quantum dots can be coherent. Blinking of the $\Xn$ emission line is much more pronounced in quasi-resonant compared to above-band excitation, as can be observed in the auto-correlation trace by the strong bunching and its decay, similarly observed elsewhere \cite{Jahn2015}. We attribute this behavior to two effects: Firstly, the fluctuating population of defect states in the matrix material environment of the QD influence its energetic structure through the quantum-confined Stark effect \cite{Miller1984}. These defect states may get saturated in the case of above-band excitation, but not in resonant pumping \cite{Besombes2003}. Secondly, the average QD charge state may get reconfigured depending on the employed excitation scheme \cite{Benny2012}. We observe that the average QD charge becomes generally more positive in quasi-resonant excitation.

In Fig. \ref{fig:basic_PL}(c) traces of cross-correlations between $\Xp$ and $\Xn$ are depicted for both above-band and quasi-resonant excitation schemes. The two correlation traces provide information with respect to the charge fluctuations between neutral and positively charged states of the QD. Again, as in the case of the $\Xn$ auto-correlation, the charge related blinking behavior is very different in the two excitation schemes. In above-band pumping the fluctuations take place on shorter timescales compared to quasi-resonant excitation. This is a further indication of the increased coherence in the quasi-resonant driving scheme, as there are far fewer free carriers in the matrix material surrounding the QD - which in turn enhances stability of the QD charge state. Under the assumption that both $\Xn$ and $\Xp$ are pumped equivalently - which is a valid assumption for above-band excitation -, the intensity ratio between $\Xn$ and $\Xp$ lines can independently be determined form the ratio $\Tcorr^{\AB}(\Xn \to \Xp)/ \Tcorr^{\AB}(\Xp \to \Xn) \simeq 4.3$. This value qualitatively matches the intensity ratio determined from the PL spectrum of Fig. \ref{fig:basic_PL}(a) of $2.8$, the difference can be attributed to the dissimilar exciton powers of $4.4$ and $1.0$ $\mu$W, respectively. Due to the presence of a double $\Xn$ and $\Xp$ resonance at $\Delta_{E_i-E_{X_S^0}} = 10.8$ meV, c.f. Fig. \ref{fig:PLE}, the assumption of equivalent excitation conditions for both lines can be extended also to the quasi-resonant excitation scheme in this particular case. The resulting intensity ratio $\Tcorr^{\QRes}(\Xn \to \Xp)/ \Tcorr^{\QRes}(\Xp \to \Xn)$ is about $0.7$ which demonstrates that the QR excitation scheme favors positive charge states in the QD. This result is a consistent observation for GaAs quantum dots, but the exact mechanism of this effect remains unclear and is beyond the scope of this work.

In Fig. \ref{fig:basic_PL}(d) the cross-correlation between the $\Xn$ and $\XmXXn$ emission lines is shown. As expected from the $\XXn \to \Xn$ cascade  a strong bunching is observed, while the $\Xn \to \XXn$ process is anti-bunched \cite{Santori2002}. Due to the nature of the cascade for each emission of a $\XXn$ photon the $\Xn$ state is inherently coherently prepared. Therefore $\Tcorr(\XXn \to \Xn)$  is equivalent to the $\Xn$ lifetime $T_1^{\Xn} = (0.303 \pm 0.006)$ ns, which matches the value $\Tcorr^{\QRes} = (0.32 \pm 0.01)$ ns. This confirms that the quasi-resonant excitation is largely coherent, at least in the case of the specific resonance used - the latter is further discussed using results for PLE and pump-probe experiments.

Fig. \ref{fig:basic_PL}(e) depicts a cross-correlation trace between $\XXXn$ and $\Xn$ emission lines recorded using above-band excitation. Analogous to the correlation trace of Fig. \ref{fig:basic_PL}(d) a clear cascade behavior - although with a lower bunching - is observed. We therefore attribute the emission line to a tri-exciton s-shell emssion line, c. f. Tab. \ref{tab:lines}.

\subsection{Excitation Spectroscopy} 

\onecolumngrid

\begin{figure}[h]
	\includegraphics[width=1.0\textwidth]{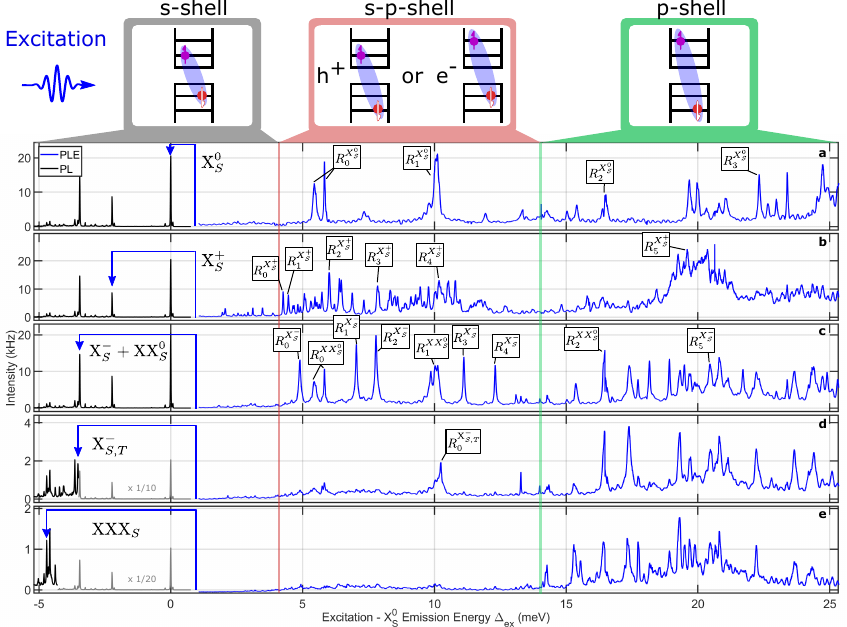}
	
	\caption{
		Combined PLE (blue) and PL (black) spectra of selected s-shell emission lines (c.f. Tab. \ref{tab:lines}) obtained by integration over s-shell lines as a function of relative energy of cw excitation laser and $\Xn$ emission energy $\Dex$. Different resonances and excitation shells are annotated, for details see Tab. \ref{tab:PLE_res} and text. Above-band spectra are recorded at $\Xn$ saturation excitation power of $1.5$ $\mu$W, while the PLE spectra are recorded by employing a $1.2$ $\mu$W quasi-resonant laser excitation.
	}
	\label{fig:PLE} 
\end{figure}

\twocolumngrid

Excited state resonances are characterized by cw PLE spectroscopy employing excitation energies relative to the $\Xn$ transition $\Dex$ in the range of $1$ to $25$ meV. The resulting PLE spectra (blue lines) of several s-shell QD emission lines are depicted in Fig. \ref{fig:PLE} as a function of $\Dex$. Reference s-shell emission spectra (black lines) obtained using above-band excitation are shown for comparison. The s-shell transitions, which are numerically integrated as a function of $\Dex$ in order to obtain the PLE spectra, are indicated by blue arrows. The spectral integration window is equivalent to the full width half maximum (FWHM) of the s-shell emission. The resolution of the PLE spectra is better than $5$ $\mu$eV, limited only by the step size of the cw tunable laser emission energy. As can readily be seen from the PLE spectra, a significant number of resonances can be found in the presented $\Dex$ range. Resonances that are relevant for the further discussion are annotated in the figure and detailed in Tab. \ref{tab:PLE_res}. For InGaAs QDs grown in the Stranski-Krastanov mode unambiguous identification of many excited state resonances has been demonstrated \cite{Bayer2002, Siebert2009, Warming2009, Benny2011a, Benny2012}. Though the same would in principle also be possible for excited states of GaAs QDs used in this study, the energetic structure of these QDs is different as outlined below. Taking into account the fact that there is presently no theory model for the energetic structure of droplet etched GaAs QDs, the rigorous identification of the PLE resonances to specific QD multi carrier excitations - beyond some general observations - is outside the scope of this work and is left to future studies.

Resonances $R_0^{X_S^0}$ and $R_0^{XX_S^0}$ of PLE spectra in Fig. \ref{fig:PLE}(a) and (c), respectively, are equivalent because every bi-excitonic photon emission is necessarily followed by emission of $\Xn$ due to the neutral exciton cascade, c. f. Fig. \ref{fig:basic_PL}(d). By using this the association between the PLE spectra of $\XmXXn$ it is possible to separate resonances of $\Xm$ and $\XXn$ which are intermingled in Fig. \ref{fig:PLE}(c). The annotated resonances reflect this association. By comparison with PLE scans performed on InGaAs QDs of Ref. \cite{Benny2012} resonances $R_0^{X_S^0}$ and $R_0^{XX_S^0}$ can be identified as the excited state configuration $(1e^1)(2h^1)$. In the study of \emph{Benny et. al.} this excited state was identified at a $\Dex$ of about $16$ meV, which is an increase of about three times compared to the GaAs QDs used in this study. It can therefore be concluded that the energetic splittings in the confinement potential of GaAs quantum dots are significantly lower. We attribute this difference to the fact that droplet etched GaAs QDs are larger in their dimensions both parallel and perpendicular to the growth direction \cite{Huber2017}. Furthermore, compared to the InGaAs QDs of Ref. \cite{Benny2012} the energetic splitting of the heavy holes in GaAs QDs is significantly lower, which can be observed in the comparatively smaller energy separation between PLE resonances of Fig. \ref{fig:PLE}(b) and (c). For example: The energetic splitting between $R_0^{X_S^+}$ and $R_1^{X_S^+}$ is $(0.20 \pm 0.02)$ meV, while for $R_0^{X_S^-}$ and $R_1^{X_S^-}$ it equates to $(2.12 \pm 0.02)$ meV, yielding a $10.7$ fold reduction of the $\Xp$ compared to $\Xm$ excited state resonance splittings. The exact mechanism causing the dissimilar energetic splittings of electronic sub-levels is currently not known, we think it could be influenced significantly by the difference of the confinement potential depths between electrons and holes.

\begin{table}[htbp!]
	
	\begin{tabularx}{\linewidth}{|Y|Y|S[table-format=2.3]|Y|S[table-format=2.0]|S[table-format=2.1]|}
			
		\hline
		{S-shell} & {Ex. Res.} & {$\Dex$ (meV)} & {Shell} & {$P_\text{sat}$ ($\mu$W)} & {$\TNR$ (ps)} \\
		\hline
		
		\multirow{6}{*}{$\Xn$} & $R_0^{\Xn}$ & 5.444 & s-p($h^+$) &  13 & 163.0 \\
		& & 5.830 & & & \\		
		& $R_1^{\Xn}$ & 10.100 & s-p($e^-$) & 8 & 218 \\
		& & & & {0.04 $P_\text{sat}$} & 36.9\\		
		& $R_2^{\Xn}$ & 16.472 & p & 14 & 389 \\		
		& $R_3^{\Xn}$ & 22.324 & p & 6 & 385\\
		\hline
		
		\multirow{7}{*}{$\Xp$} & $R_0^{\Xp}$ & 4.264 & s-p($h^+$) & &\\		
		& $R_1^{\Xp}$ & 4.464 & s-p($h^+$) & &\\		
		& $R_2^{\Xp}$ & 6.017 & s-p($h^+$) & 6 & 129\\
		& & & & {0.03 $P_\text{sat}$} & 88.3\\
		& $R_3^{\Xp}$ & 7.781 & s-p($h^+$) &  & \\		
		& $R_4^{\Xp}$ & 10.179 & s-p($e^-$) & 9 & 210\\		
		& $R_5^{\Xp}$ & 19.591 & p & 1 & 165\\
		\hline
		
		\multirow{7}{*}{$\Xm$} & $R_0^{\Xm}$ & 4.900 & s-p($h^+$) & 10 & 56.8 \\
		& & & & {0.04 $P_\text{sat}$} & 29.1\\				
		& $R_1^{\Xm}$ & 7.040 & s-p($h^+$) & & \\		
		& $R_2^{\Xm}$ & 7.786 & s-p($h^+$) & 20 & 74.1 \\		
		& $R_3^{\Xm}$ & 11.119 & s-p($h^+$) & & \\		
		& $R_4^{\Xm}$ & 12.305 & s-p($h^+$) & & \\		
		& $R_5^{\Xm}$ & 20.464 & p & 4 & 240 \\
		\hline
		
		\multirow{4}{*}{$\XXn$} & \multirow{2}{*}{$R_0^{\XXn}$} & 5.444 & \multirow{2}{*}{s-p($h^+$)} & & \\		
		& & 5.834 & & &\\		
		& $R_1^{\XXn}$ & 10.027 & s-p($e^-$) & &\\		
		& $R_2^{\XXn}$ & 16.469 & p & & \\
		\hline
		
	\end{tabularx}		
	\caption{List of annotated excitation resonances of Fig. \ref{fig:PLE}(a-c). Excitation resonance energies $\Dex$ given relative to the s-shell $\Xn$ transition. The accuracy of $\Dex$ is estimated to $16$ $\mu$eV. Each excitation resonance is attributed to a specific series of transitions, called shells. Details and definitions are found in the text. Excitation saturation powers $P_\text{sat}$ and non-radiative decay times $\TNR$ obtained by modeling of pump-probe experiments, c.f. Fig. \ref{fig:PumpProbe} and Suppl. Fig. \ref{fig:PumpProbe_Full}, are shown for selected resonances. The statistical standard deviation of $\TNR$ is about $1$ $\%$ at $P_\text{sat}$ and to about $5$ $\%$ at 0.04 $P_\text{sat}$.} 
	\label{tab:PLE_res}
\end{table}

\normalsize

While the exact identification of most excited state resonances is beyond the scope of this study, it is clear that there should be different series of excitation resonances - very much analogous to the spectral series (Lyman, Balmer, Paschen, etc.) in hydrogen atoms \cite{Bohr1954} - present in this system. Just as in the case of hydrogen atoms the excitation resonance with the lowest $\Dex$ can be considered the fundamental transition, while the series of excitation resonances extends towards increasing $\Dex$. In contrast to atoms in QDs the individual transitions of a series originate not from higher order electronic orbitals but from transitions between different excitonic (spin) complexes. In order to accommodate this difference the series of excitation state resonances in QDs are henceforth called shells. The relevant ones for the discussion of the PLE spectra of Fig. \ref{fig:PLE} are: s-shell [$\ket{0} \to (1e^1)(1h^1)$], s-p-shell($h^+$) [$\to (1e^1)(2h^1)$], s-p-shell($e^-$) [$\to (2e^1)(1h^1)$] and p-shell [$\to (2e^1)(2h^1)$], defined in accordance to \cite{Benny2012}. The fundamental transitions of each shell are presented schematically at the top of Fig. \ref{fig:PLE}. Since the PLE trace of Fig. \ref{fig:PLE} was recorded at about one third of the typical saturation intensity of most transitions of about $4$ $\mu$W, it can be concluded that the relative intensities of the different excitation resonances generally reflect the strength of the light-matter interaction matrix element of the individual transitions. Due to the nature of the different energetic orbitals of electronic states in QDs the dipole moment between individual series of excitation resonances can be significantly different \cite{Warming2009, Benny2011a}. At the boundary between s- and s-p-shell resonances we therefore expect to detect a significant change in the intensity of the observed basic s-shell transitions (Fig. \ref{fig:PLE}(a-c)). Consequently, we attribute the first bright PLE resonances to the lower boundary of the s-p-shell of $\Dex \simeq 4.2$ meV. This boundary is indicated in Fig. \ref{fig:PLE} by a red line.

In Fig. \ref{fig:PLE}(d) the PLE spectrum of a spin-blockaded negative trion ($\XmT$) is shown. This state is a negatively charged exciton in which the electron spin configuration is such that a single electron spin is locked in the p-shell by the Fermi exclusion principle due to equivalent spins of s- and p-shell electrons. The details and properties of this state are discussed elsewhere \cite{Benny2014}. Due to the presence of an electron in the p-shell the s-p-shell($h^+$) transitions are blocked, while the ones of the electron, i.e. the s-p-shell($e^-$), are still allowed. The consequence can be seen in the PLE spectrum: most resonances $\Dex < 14$ meV that are observed in Fig. \ref{fig:PLE}(a-c) are absent, which also confirms the identity of the observed s-shell emission line. We therefore conclude that the absent excitation resonances belong to the s-p-shell($h^+$), while $R_0^{\XmT}$ belongs to the s-p-shell($e^-$). 

Fig. \ref{fig:PLE}(e) depicts the PLE spectrum of a s-shell line we associate with the s-shell transition of a tri-excitonic complex [$(1e^2 2e^1)(1h^2 2h^1)$]. As a consequence both s-p-shells are absent in its PLE spectrum. The first excitation resonance can therefore be found in the p-shell, which enables the estimation of the lower p-shell boundary to $\Dex \simeq 14.1$ meV. This is indicated in Fig. \ref{fig:PLE} by a green line. Around $\Dex \simeq 20$ meV a broad maximum in the PLE spectra, especially in the one of $\Xp$, is observed. This can be attributed to the overlap of p-shell transitions of both electrons and holes and not to effects related due to phonon-enhanced absorption - which would be expected beyond $\Dex \gtrsim 32$ meV \cite{Adachi1993, Benny2012}.

\subsection{Pump-Probe Correlation Spectroscopy}

\begin{figure}[h]
	\includegraphics[width=1.0\linewidth]{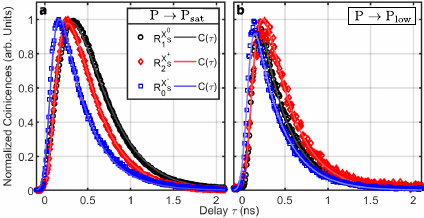}
	
	\caption{\label{fig:PumpProbe} 
		Quasi-resonant pump-probe correlation measurements of $\Xn$, $\Xp$ and $\Xm$ fundamental QD transitions and their respective excitation resonances (a) at excitation saturation and (b) far below saturation power. The curves are modeled according to Eq. (\ref{eq:pump_probe}) with $\TNR$ as the only free parameter. Specific resonances, excitation conditions and estimated $\TNR$ values are summarized in Tab. \ref{tab:lines}. }
\end{figure}

In order to investigate the non-radiative decay mechanism of the excitation resonances, pump-probe experiments are conducted. In these investigations specific excitation resonances are pumped by a pulsed laser while the time resolved emission of the respective s-shell transition is monitored. The results for selected $\Xn$, $\Xp$ and $\Xm$ transitions and resonances pumped at and far below ($\approx 0.04 P_\text{sat}$) excitation saturation are shown in Fig. \ref{fig:PumpProbe}(a) and (b), respectively. Further results of pump-probe experiments are plotted in Suppl. Fig. \ref{fig:PumpProbe_Full}. The intensity correlation data is modeled by a delayed exponential decay function

\begin{align}
\label{eq:pump_probe}
C(\tau) & = e^{-\frac{\tau }{T_1^{\Xn}}} \theta (\tau ) \ast \left( \frac{e^{-\frac{\tau }{\TNR}}}{\TNR} \ast \DetResp \right)  \;,
\end{align}

where the non-radiative decay time $\TNR$ is the only free parameter. All other parameters are fixed to values determined previously. The delay which is induced by a single particle non-radiative decay process is modeled by an exponential distribution $\frac{e^{-\frac{\tau }{\TNR}}}{\TNR}$. This simple assumption fits the experimental data very well for most resonances, for the curve shown in Suppl. Fig. \ref{fig:PumpProbe_Full}(c) - the model shows deviations to the experimental data. This can be attributed to the break down of the assumption of a single stage decay process. A summary of the extracted $\TNR$ values can be found in Tab. \ref{tab:PLE_res}. For all basic QD charge states the decay time increases towards elevated values of $\Dex$ and towards excitation power saturation. As a consequence, the observed excitation resonances with the lowest $\TNR$  - and therefore the most coherent excitation process - can be found in the s-p-shells. Note that no $\pi$-pulses, a sign of coherent preparation of the QD s-shell states, are observed for any excitation resonance \cite{Kamada2001, Fox2006}, the reason is that the required condition $\TNR \ll T_1^{\Xn}$ of the optical Bloch equations is not fulfilled at saturation of the excitation resonances. Far below saturation, where the influence of the excitation laser induced increase of the decay time should be neglectable, the $\TNR$ values of $(37 \pm 2)$, $(88 \pm 2)$ and $(29 \pm 1)$ ps are determined for the QD s-shell transitions $\Xn$, $\Xp$ and $\Xm$, respectively. The excitation resonances in which an additional electron is present in the QD exhibit the fastest non-radiative decay times, while if an additional hole is present the decay is about a factor of $3$ slower. Consequently, the value of $\TNR$ in the absence of additional carriers lies between the two former cases. By considering the natural spectral line broadening induced by the lifetime of the excited states limited by $\TNR$, these differences explain the variation of the observed spectral line widths of the PLE resonances of Fig. \ref{fig:PLE}.

\subsection{Polarization Resolved Emission Spectroscopy}

\begin{figure}
	\includegraphics[width=1.0\linewidth]{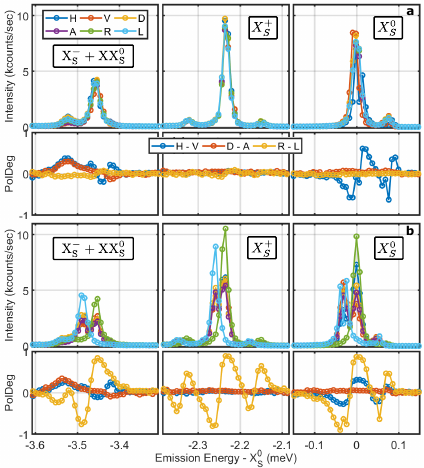}
	
	\caption{\label{fig:PolPL} 
		Polarization resolved PL spectra of $\XmXXn$, $\Xp$ and $\Xn$ without applied external magnetic field (a) and with a field of $0.2$ T (b), respectively. The spectra are recorded using above-band excitation at $P \simeq P_{sat}/2$. The raw spectra are shown in the upper panels, while in the lower ones the degree of polarization (PolDeg, see text) between orthogonal bases are depicted.}
\end{figure}

In order to characterize the spin related properties of excitonic complexes in QDs, a convenient method is to investigate the polarization characteristics of the emission. This basic characterization is shown with and without applied magnetic field in Fig. \ref{fig:PolPL}. To distinguish the spectral polarization dependence the degree of polarization is defined as $\PolDeg(\text{I-J}) = \frac{\rho(I)-\rho(J)}{\rho(I)+\rho(J)}$, where $I$ and $J$ represent a set of orthogonal polarization bases and $\rho$ are the spectral intensities associated with the respective bases.  In the absence of an external magnetic field, $\Xn$ features a finestructure splitting due to its integer quasi-particle spin $j_{\Xn} = \pm1$ which is oriented along the H-V polarization axis, cf. Fig. \ref{fig:PolPL}(a). For finite magnetic fields, depending on the field strength, the eigenstates are oriented on a superposition axes of H-V and R-L (c.f. Fig. \ref{fig:PolPL}(b)), and are split according to the Zeeman splitting \cite{Bayer2002}. $\Xp$ does not exhibit a fine structure in the absence of a magnetic field (${j_{\Xp} = \pm 3/2, j_{\Xp} = \pm 1/2}$) due to the Kramers degeneracy of half-integer spin states \cite{Kramers1930, Bayer2002}. Its polarization eigenbase is therefore R-L. As the excitation is not polarization selective it does not show any polarization dependence in Fig. \ref{fig:PolPL}(a). In the presence of a magnetic field $\Xp$ exhibits Zeeman splitting, c.f. Fig. \ref{fig:PolPL}(b), which allows the QD g-factor of $1.86 \pm 0.02$ to be determined. Due to the overlap of the $\XmXXn$ transitions, the signature of the $\XXn$ finestructure splitting in the polarization resolved spectra is masked.


The polarization resolved emission spectra of Fig. \ref{fig:PolPL} are used to calibrate the polarization axes of the QD eigenbases to the lab frame. In this process the fundamental polarization quantization axes are used in the following way: The finestructure splitting of $\Xn$ is used to calibrate the D-A and R-L axes to an orthogonal orientation with respect to H-V. The eigenbases shift to R-L induced by a small magnetic field is used to minimize the polarization degree in D-A axis. As the orthogonality between the axes is preserved, this fully defines the rotation between the QD and lab frames. The degree of polarization of Fig. \ref{fig:PolPL} shows that this calibration is not fully equivalent for all investigated transitions. This effect can be attributed to the wavelength dependent birefringent behavior of the employed GaP microlenses (see methods section) - as this relation is not present in samples without these lenses.

\subsection{Polarization Resolved Excitation Spectroscopy}
\label{sec:pol_PLE}


\begin{figure}[h]
	\includegraphics[width=1.0\linewidth]{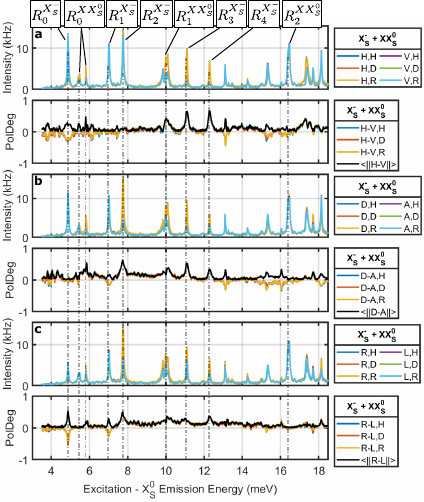}
	
	\caption{\label{fig:PolPLE} 
		Exemplary polarization resolved PLE spectra and polarization degrees (PolDeg) in 18 different combinations of excitation and detection polarization bases (I,J) of the $\XmXXn$ emission lines. The data corresponds to the third run of Fig. \ref{fig:AbsPolDeg}. The first (I) and second (J) bases are of the excitation and detection, respectively. Upper panels: Combinations with excitation polarization bases (a) H/V, (b) D/A and (c) R/L. The respective excitation PolDegs are shown in the lower panels. Additionally, total PolDegs of each excitation axes $\ew{\lVert \text{I-J} \rVert}$ according to Eq. (\ref{eq:ex_poldeg}) are drawn. Resonances corresponding to Tab. \ref{tab:PLE_res} are annotated.}
\end{figure}

Polarization resolved PLE spectra are depicted exemplarily in Fig. \ref{fig:PolPLE} for the $\XmXXn$ transitions. Polarization resolved PLE spectra of $\Xn$ and $\Xp$ emission lines are found in Suppl. Figs. \ref{fig:PolPLE_Xn} and \ref{fig:PolPLE_Xp}, respectively. As can readily be observed, the behavior of the various resonances is not equivalent, c.f. $R_1^{\Xm}$ vs. $R_2^{\Xm}$ vs. $R_3^{\Xm}$. There are two reasons for this, both of which can be negated similarly to the polarization resolved PL by calibration of the excitation polarization bases to the spin eigenstates of the specific excitation resonance. Firstly, the polarization calibration of the excitation changes significantly with wavelength due to the birefringent behavior of the GaP microlens. Secondly, the excitation resonances origin from different excitonic complexes with distinct spin configurations. Depending on whether the excited and ground state spin symmetries match and if the non-radiative decay process preserves the initial spin created in the excited state, the polarization of the QD excitation and emission are correlated. In order to extract this correlation, which constitutes the s-shell spin preparation fidelity, in the presence of the birefringent GaP microlens the PLE spectra of Fig. \ref{fig:PolPLE} are measured in 18 different combinations of polarization bases, i.e. 6 excitation and 3 detection bases. To avoid calibrating each excitation resonance separately, the excitation polarization degree $\ew{\lVert \text{I-J} \rVert}$ may be defined as

\begin{align}
\label{eq:ex_poldeg}
 \text{PolDeg}_{\text{ex}}(\text{I-J})  & :=  \ew{\lVert \text{I-J} \rVert} \\
 & = \sqrt{\frac{\sum _K \left. (\rho(\text{I})-\rho(\text{J}))^2\right\vert_{\text{K}}}{3 \; \sum _K \left. (\rho(\text{I})+\rho(\text{J}))^2\right\vert_{\text{K}}}} \; ,
\end{align}    

where K is the observant (detection) polarization base. Exemplary results of $\ew{\lVert \text{I-J} \rVert}$ are shown in the lower panels of Fig.  \ref{fig:PolPLE} together with the individual polarization degrees $\left. \text{PolDeg}(\text{I-J})\right\vert_{\text{K}}$. It can be observed that $\ew{\lVert \text{I-J} \rVert}$ effectively constitutes the vector norm of the polarized response of the QD via the excitation resonance. In other words it renders the calibration of specific excitation resonances to the fundamental QD eigenstates unnecessary.

In order to determine the total spin preparation fidelity of the excitation resonances the polarization degree can be abstracted further. This can be done irrespective of varying detection and excitation polarization bases calibration in respect to specific eigenstates and excitation resonances. The total excitation polarization degree, which is equivalent to the ground state spin preparation fidelity $f_\text{SpinPrep}$, is therefore defined as

\begin{align}
\label{eq:tot_ex_poldeg}
f_\text{SpinPrep} & \equiv \lVert \text{PolDeg}_{\text{ex}} \rVert  := \sqrt{\sum _{\text{I-J}} \ew{\lVert \text{I-J} \rVert}^2}  \; .
\end{align}  

\begin{figure}
	\includegraphics[width=1.0\linewidth]{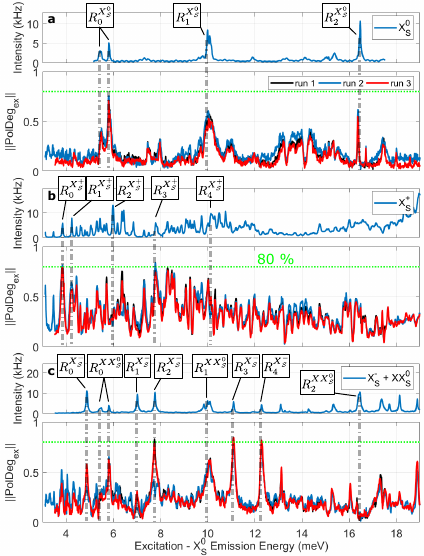}
	
	\caption{\label{fig:AbsPolDeg} 
		Spin preparation fidelity $\SpinPrep \equiv \lVert \text{PolDeg}_{\text{ex}} \rVert$ (c.f. Eq. (\ref{eq:tot_ex_poldeg})) spectra for QD s-shell transitions $\Xn$, $\Xp$, $\XmXXn$ and multiple experimental runs, more details are given in the text.  Resonances of Tab. \ref{tab:PLE_res} are annotated by labels and gray dashed guidelines. The $80$ $\%$ spin preparation fidelity threshold is indicated in green. }
\end{figure}

Spectra of $\SpinPrep$, extracted from the polarization resolved PLE spectra in analogue to Fig. \ref{fig:PolPLE} of $\Xn$, $\Xp$ and $\Xm + \XXn$, are summarized in Fig. \ref{fig:AbsPolDeg}.

By employing this method the spin memory of different excitation resonances can be compared effectively. For example the polarization response of resonances $R_2^{\Xn}$ and $R_2^{\XXn}$ does not show any dependence on the excitation polarization, indicating that the excited and ground states exhibit orthogonal eigenstates. Resonances $R_0^{\Xp}$, $R_3^{\Xp}$, $R_3^{\Xm}$ and $R_4^{\Xm}$ on the other hand show $\SpinPrep$ values up to $85$ $\%$, demonstrating matching excited and ground spin configurations as well as spin preserving non-radiative decay processes. For $\Xn$ the excitation resonance induced spin memory is limited to $75$ $\%$, c.f. $R_0^{\Xn}$. Other resonances show intermediate spin memory effects, which can be attributed to either partial mismatch between excited and ground state eigenstates or non-polarization preserving non-radiative decay processes. A reason for the latter would be, if more than one particle is involved in the decay process. The resonances with the highest spin preparation fidelities all belong to the s-p-shell, c.f. Fig. \ref{fig:PLE}, indicating that resonances of this shell would be preferable for implementations of quantum spin memories based on droplet etched GaAs QDs.

Polarization resolved PLE spectra are recorded in three separate experimental runs which feature different experimental conditions. The first, second and third runs are optimized for detection polarization bases of $\Xn$, $\Xp$ and $\Xm$, respectively. Additionally, weak above-band excitation ($<2$ nW) is employed and the excitation polarization bases are aligned using the reflection of a resonant laser in the first and third runs, while no above-band excitation and excitation calibration to $R_1^{\Xp}$ is used in the second. The resonant excitation powers of the three runs are $1.4$, $4$ and $4$ $\mu$W, respectively. Generally the results between different experimental runs are very consistent - underling the validity and robustness of the method of extraction of $f_\text{SpinPrep}$ presented in this study. Only the resonance $R_2^{\Xm}$ shows significant deviation between the experimental runs, indicating the influence of above-band excitation for this specific excitation resonance.

\section{Conclusions}

For the first time droplet etched GaAs QDs are investigated comprehensively using polarization resolved PL, PLE and correlation spectroscopy. It is found that the GaAs QDs carrier dynamics depend drastically on the excitation method, e.g. in continuous wave quasi-resonant excitation blinking of the $\Xn$ transition is very pronounced. Consequently, the  on-off ratio of $\Xn$ in above-band excitation is about $8$ times higher. We attribute this behavior to the defect states around the QDs, similarly reported in Ref. \cite{Jahn2015}. Furthermore, the predominant charge of the QDs is shifted towards excitations with excess holes in quasi-resonant excitation schemes. The lifetime of the $\Xn$ transition is determined consistently using both quasi-resonant and above-band correlation spectroscopy to $(303 \pm 6)$ ps. Using this information, the non-radiative decay processes of several different excitation resonances for $\Xn$, $\Xp$ and $\Xm$ transitions are investigated by pump-probe experiments. The minimal characteristic non-radiative decay times $\TNR$ are estimated to $37$, $88$ and $29$ ps, respectively, all of which can be attributed to s-p-shell excitation resonances. $\TNR$ is found to increase for excitation powers as well as higher energy (e.g. p-shell) resonances. The energetic structure of the excitation resonances is investigated using PLE spectroscopy. By employing observed properties of various excitonic complexes specific resonances can be attributed to different energetic shells. In order to match the polarization eigenbases of the laboratory frame to the one of the QDs, its polarization eigenstates both with and without external magnetic field are employed. Due to the wavelength dependent birefringency of the GaP microlenes, used to enhance the collection efficiency of the QD emission \cite{Chen2018}, the polarization calibration is wavelength dependent. To effectively extract the spin preparation fidelity $\SpinPrep$ independently of the calibration of excitation and detection polarization bases a novel method is proposed. This procedure employs measurements in $18$ different excitation and detection bases and the orthogonality between bases pairs to determine $\SpinPrep$ obtained using excitation resonances. $\SpinPrep$ of up to $85$ $\%$ for both $\Xp$ and $\Xm$ as well as about $75$ $\%$ for $\Xn$ are achieved for s-p-shell type transitions. 

In conclusion the presented comprehensive investigations, methods and findings enable the directed usage of excited state resonances to deterministically prepare fundamental spin states in GaAs quantum dots. These fundamental abilities will pave the way to use these quantum dots in a large variety of future quantum optical experiments and applications, such as QD based quantum memories \cite{Stockill2016,Cogan2018,Chekhovich2020}, quantum entanglement repeaters \cite{Zopf2019,BassoBasset2019}, photon graph \cite{Wang2016} and cluster states \cite{Schwartz2016} as well as more efficient entangled photon pair sources \cite{Chen2018, Hopfmann2020}. Additionally, the presented study increases the detailed understanding of droplet etched GaAs QDs significantly. While not all aspects are explored exhaustively, it provides an ideal starting point for more detailed investigations in a variety of aspects. Examples of the latter are identification of transitions of excitonic complexes, excitation scheme dependent photonic and electronic coherences, the nature of the non-radiative decay process and dependency of QD properties on matrix material composition.




\section{Methods}

\begin{figure}
	\includegraphics[width=1.0\linewidth]{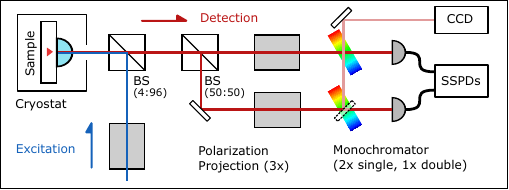}
	
	\caption{\label{fig:Setup} 
		Simplified schematic drawing of the experimental setup. The following abbreviations are used: Beam splitter (BS), superconducting single photon nano wire detectors (SSPDs) and charge coupled device (CCD).}
\end{figure}

The presented study is performed using droplet etched GaAs quantum dots (QDs) embedded in QD-nanomembranes and attached to GaP microlenes. Samples were grown on [001] GaAs substrate using molecular beam epitaxy. The nanoholes are etched through deposition of Al droplets onto the $\AlGaAs$ matrix material. Consequently, the $15$ nm deep and $40$ nm wide nanoholes are filled with GaAs thereby forming the QDs. The $380$ nm thick $\AlGaAs$ matrix material with the centrally embedded QDs is grown on-top of a AlAs sacrificial layer, which is removed by selective hydrofluoric acid etching to yield the QD-nanomembranes. In order to overcome the strong internal reflection of the semiconductor material due to its high refractive index ($\simeq 3.5$) and enhance the out-coupling efficiency the QD-nanomembranes are attached to GaP microlenses using a $50$ nm thick layer of PMMA. This enhances the photon extraction efficiency compared to unprocessed samples by a factor of about $100$. A detailed description of the growth and processing steps can be found in Refs. \cite{Wang2007, Chen2018}.\\

The employed experimental setup is shown as a simplified sketch in Fig. \ref{fig:Setup}. The QD-devices consisting of GaP microlenses with attached QD-nanomembranes are placed inside a dry closed-cycle cryostat. The cryogenic system features a z-axis $9$ T superconducting magnet and a $3.8$ K base temperature. Sample luminescence is collected using a aspheric lens of $0.64$ NA. Photoluminescence (PL) spectroscopy is performed using a $2 \times 0.750$ m double spectrometer and gratings of either $1800$ or $1200$ lines/mm. The maximal spectral resolution of this system is about $15$ $\mu$eV at $780$ nm. The spectrometer can be configured as two independent monochromators, which is employed in cross-correlation experiments. Spectroscopic investigations are performed by a standard back-illuminated deep-depletion CCD. For PLE spectroscopy a narrow-band ($100$ MHz) wavelength-tuneable and -stabilized cw laser, in conjunction with the high resolution PL detection system, is employed. Wavelengths are tuned and stabilized to an absolute accuracy of $2$ pm using a calibrated wavelength meter. In order to enhance the suppression of the excitation laser and separate it from the QD emission tuneable band edge filters in both excitation and detection are employed. Pump-probe experiments are performed using a wavelength tuneable and pulsed optical parametric oscillator (OPO) laser pumped by a pulsed frequency doubled fiber laser at $516$ nm. The OPO laser system exhibits a repetition rate of  $76.271$ MHz and a shaped pulse width of $3.5$ ps. For above-band excitation a HeNe laser featuring a $632.8$ nm emission wavelength is employed. In order to perform polarization resolved PL and PLE spectroscopy the polarization of the QD emission is projected onto one of the set of orthogonal polarization bases [H, V, D, A, R, L]. This is achieved by using two liquid crystal variable retarders (LCVRs) and a linear polarizer. Aforementioned configuration can be calibrated to any orthogonal set of polarization bases on the Poincaré sphere and enables the compensation of the mismatch between QD and lab polarization eigenbases. The accuracy of LCVR calibration was determined to about $0.95$ using a polarimeter. Three independent polarization projector units are employed: two in the detection arms and one in the excitation path, c.f. Fig. \ref{fig:Setup}. Correlation spectroscopy is performed by time resolved correlation of electronic signals from superconducting nanowire single photon detectors (SSPDs). The time resolution of this system is about $53$ ps for one and $75$ ps for two-photon correlations.\\


\begin{acknowledgments}
	We thank Michael Zopf and Jingzhong Yang (LU Hannover) for fruitful discussions. We acknowledge funding by the BMBF (Q.link.X) and the European Research Council (QD-NOMS).
\end{acknowledgments}

\section{Supplementary}

\renewcommand{\theequation}{S\arabic{equation}}
\renewcommand{\thefigure}{S\arabic{figure}}

\subsection{Fine structure}

\begin{figure}[htp]
	\includegraphics[width=1.0\linewidth]{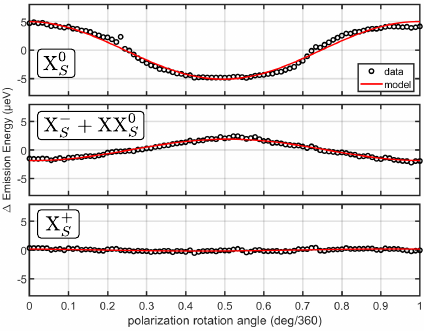}
	
	\caption{\label{fig:Sup_FSS} 
		Relative energies of selected emission lines ($\Xn$, $\Xp$, $\XmXXn$) as a function of the polarization projection angle. Modeling to simple sine functions (see text) are shown in red. }
\end{figure}

Fig. \ref{fig:Sup_FSS}(a-c) depicts the relative emission energies of $\Xn$, ($\XmXXn$) and $\Xp$, respectively, as a function of the polarization rotation angle. While $\Xn$ as well as $\XmXXn$ exhibit an anti-correlated interdependence signaling the presence of the neural exciton finestructure splitting, $\Xp$ does not. The latter is a consequence of Kramers degeneracy of half-integer spin states \cite{Kramers1930, Bayer2002}.  The data is modeled using sinusoidal functions: $\Delta E(\alpha) = E_0 + \Delta_{X^i}^{FSS}/2 \: \text{sin}(2 \pi \alpha + \alpha^0)$, where the two free parameters are the fine structure splitting $\Delta_{X^i}^{FSS}$ and phase $\alpha^0$. The model yields a fine structure splitting of $\Delta_{\Xn}^{FSS} = (10.1 \pm 0.3)$ $\mu$eV for $\Xn$, while there is no significant splitting for $\Xp$. Since the lines $\Xm$ and $\XXn$ cannot be spectrally separated a reduced fine structure splitting $\Delta_{\XXn}^{FSS} = (3.8 \pm 0.1)$ $\mu$eV compared to $\Xn$ is observed. Note that $\alpha^0_{\Xn} = 0.75 \pm 0.06$ and $\alpha^0_{\XmXXn} = 0.26 \pm 0.06$ have, as expected, orthogonal phases. These observations also support the line identifications of Tab. \ref{tab:lines}.

\subsection{Pump-Probe Correlation Spectroscopy}

\begin{figure}[htp]
	\includegraphics[width=1.0\linewidth]{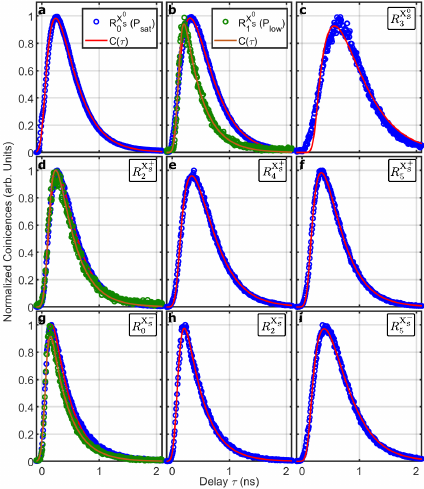}
	
	\caption{\label{fig:PumpProbe_Full} 
		Quasi-resonant pump-probe correlation measurements using different QD charge states and their respective excitation resonances. The curves are modeled according to Eq. (\ref{eq:pump_probe}) with $\TNR$ as the only free parameter. Specific resonances, corresponding to Tab. \ref{tab:PLE_res}, are annotated in the graphs. }
\end{figure}

In Fig. \ref{fig:PumpProbe} pump-probe experiments employing quasi-resonant excitation resonances $R_1^{\Xn}$, $R_2^{\Xp}$ and $R_0^{\Xm}$ are shown for both saturation and low excitation powers. Pump-probe investigations of additional resonances are depicted in Fig. \ref{fig:PumpProbe_Full}. These correlation traces are modeled according to Eq. (\ref{eq:pump_probe}), the corresponding extracted characteristic non-radiative decay times are summarized in Tab. \ref{tab:PLE_res}. Except $R_3^{\Xn}$ of Fig. \ref{fig:PumpProbe_Full}(c) all obtained correlation traces fit very well to the simple model provided by Eq. (\ref{eq:pump_probe}). Since the model assumes a single stage decay process, it follows that most non-radiative decay processes involve a single process - with $R_3^{\Xn}$ as an exception. Interestingly, a strong dependence of $\TNR$ on the excitation power is only observed for $R_1^{\Xn}$, $R_2^{\Xp}$ and $R_0^{\Xm}$ are affected to a reduced extend. The origin of this effect is currently not known but could be a topic of future investigations.\\

\subsection{Polarization Resolved Excitation Spectroscopy}


\begin{figure}[htp]
	\includegraphics[width=1.0\linewidth]{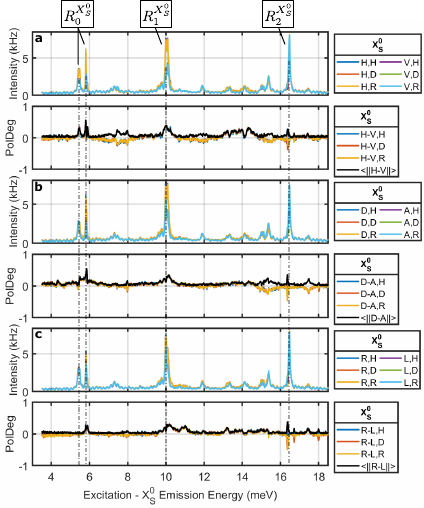}
	
	\caption{\label{fig:PolPLE_Xn} 
		Polarization resolved PLE spectra and polarization degrees (PolDeg) in 18 different combinations of excitation and detection polarization bases (I,J) of the $\Xn$ emission line. The data corresponds to the third run of Fig. \ref{fig:AbsPolDeg}. The first (I) and second (J) bases are of the excitation and detection, respectively. Upper panels: Combinations with excitation polarization bases (a) H/V, (b) D/A and (c) R/L. The respective excitation PolDegs are shown in the lower panels. Additionally, total PolDegs of each excitation axes $\ew{\lVert \text{I-J} \rVert}$ according to Eq. (\ref{eq:ex_poldeg}) are drawn. Resonances corresponding to Tab. \ref{tab:PLE_res} are annotated.}
\end{figure}

\begin{figure}[htp]
	\includegraphics[width=1.0\linewidth]{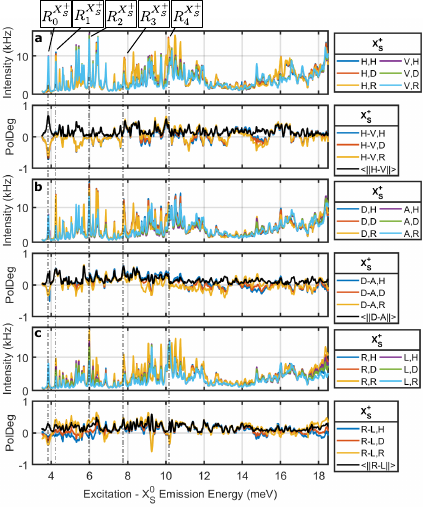}
	
	\caption{\label{fig:PolPLE_Xp} 
		Polarization resolved PLE spectra and polarization degrees (PolDeg) in 18 different combinations of excitation and detection polarization bases (I,J) of the $\Xp$ emission line. The data corresponds to the third run of Fig. \ref{fig:AbsPolDeg}. The first (I) and second (J) bases are of the excitation and detection, respectively. Upper panels: Combinations with excitation polarization bases (a) H/V, (b) D/A and (c) R/L. The respective excitation PolDegs are shown in the lower panels. Additionally, total PolDegs of each excitation axes $\ew{\lVert \text{I-J} \rVert}$ according to Eq. (\ref{eq:ex_poldeg}) are drawn. Resonances corresponding to Tab. \ref{tab:PLE_res} are annotated.}
\end{figure}

Polarization resolved PLE spectroscopy of $\XmXXn$ is shown in Fig. \ref{fig:PolPLE} for the conditions of run $3$ discussed in Sec. \ref{sec:pol_PLE} of the main text. Corresponding PLE spectra for $\Xn$ and $\Xp$ transitions are depicted in Fig. \ref{fig:PolPLE_Xn} and \ref{fig:PolPLE_Xp}, respectively. The spin preparation fidelity plotted in Fig. \ref{fig:AbsPolDeg} is determined numerically according to Eqs. (\ref{eq:ex_poldeg}-\ref{eq:tot_ex_poldeg}) using data of all three polarization resolved PLE spectra. Resolved PLE traces relevant excitation resonances of Tab. \ref{tab:lines} and Fig. \ref{fig:PLE} are annotated by $R^\text{X}_i$. It can be observed that the polarization resolved response - including its polarization Eigenbase - is very much dependent on the specific excitation resonance.. in question. \\

\pagebreak [4]

\bibliography{bibliography}

\end{document}